\documentclass[a4paper]{jpconf}
\usepackage{graphicx}
\usepackage{amssymb,amsmath,cite}

\newcommand{\ba}{\begin{eqnarray}}
\newcommand{\ea}{\end{eqnarray}}
\newcommand{\ban}{\begin{eqnarray*}}
\newcommand{\ean}{\end{eqnarray*}}
\newcommand{\bsub}{\begin{subequations}}
\newcommand{\esub}{\end{subequations}}
\begin{document}
\title{Partial dynamical symmetries in quantum systems}

\author{A Leviatan}

\address{Racah Institute of Physics, The Hebrew University, 
Jerusalem 91904, Israel}

\ead{ami@phys.huji.ac.il}

\begin{abstract}
We discuss the the notion of a partial dynamical symmetry (PDS), for which 
a prescribed symmetry is obeyed by only a subset of solvable eigenstates, 
while other eigenstates are strongly mixed. 
We present an explicit construction of Hamiltonians with this property, 
including higher-order terms, and portray their significance  
for spectroscopy and shape-phase transitions in nuclei.
The occurrence of both a single PDS, relevant to stable structures, and 
of several PDSs, relevant to coexistence phenomena, are considered.
\end{abstract}
\section{Introduction}
\label{sec:intro}

Models based on spectrum generating algebras form a convenient framework
to examine~underlying symmetries in dynamical systems, 
and have been used extensively in diverse areas of physics~\cite{BNB}. 
Notable examples in nuclear physics are 
Wigner's spin-isospin SU(4) supermultiplets~\cite{WIG}, 
SU(2) single-$j$ pairing~\cite{Kerman61}, 
Elliott's SU(3) model~\cite{Elliott58}, symplectic model~\cite{Rowe85}, 
Ginocchio's monopole and quadrupole pairing models~\cite{GIN}, 
interacting boson models (IBM) for
even-even nuclei~\cite{ibm} and boson-fermion models (IBFM) for 
odd-mass nuclei~\cite{ibfm}. 
Similar algebraic techniques have proven to be useful in the 
structure of molecules~\cite{vibron,Frank94} and of hadrons~\cite{BIL}. 
In such models the Hamiltonian is expanded in elements 
of a Lie algebra, ($G_0$), 
called the spectrum generating algebra. 
A dynamical symmetry occurs if the Hamiltonian
can be written in terms of the Casimir operators 
of a chain of nested algebras, 
$G_0\supset G_1 \supset \ldots \supset G_n$~\cite{Iachello06}. 
The following properties are then observed. 
(i)~All states are solvable and analytic expressions
are available for energies and other observables. 
(ii)~All states are classified by quantum numbers, 
$\vert\alpha_0,\alpha_1,\ldots,\alpha_n\rangle$, 
which are the labels of the irreducible representations (irreps) of the 
algebras in the chain. 
(iii)~The structure of wave functions is completely dictated by symmetry
and is independent of the Hamiltonian's parameters. 

A~dynamical symmetry provides clarifying insights 
into complex dynamics and its 
merits are self-evident. 
However, in most applications to realistic systems,
the predictions of an exact dynamical symmetry are rarely fulfilled 
and one is compelled to break it. 
The breaking of the symmetry is required for a number of reasons. 
First, one often finds that the assumed symmetry is 
not obeyed uniformly, {\it i.e.}, 
is fulfilled 
by only some of the states but not by others. Certain degeneracies
implied by the assumed symmetry are not always realized, 
({\it e.g.}, axially deformed nuclei rarely fulfill the IBM SU(3) 
requirement of degenerate $\beta$ and $\gamma$ 
bands~\cite{ibm}). Secondly, forcing the Hamiltonian to be 
invariant under a symmetry group may impose constraints which are too severe
and incompatible with well-known features of 
the dynamics ({\it e.g.}, the models
of~\cite{GIN} require degenerate single-nucleon energies). 
Thirdly, in describing systems 
in-between two different structural phases,  
{\it e.g.}, spherical and deformed nuclei, the Hamiltonian 
by necessity mixes terms with different symmetry character. 
In the models mentioned above, the required symmetry breaking is achieved
by including in the Hamiltonian terms associated with (two or more) 
different sub-algebra chains of the parent spectrum generating algebra. 
In general, under such circumstances, solvability is lost,
there are no remaining non-trivial conserved quantum numbers and all
eigenstates are expected to be mixed.
A partial dynamical symmetry (PDS)~\cite{lev11} corresponds to
a particular symmetry breaking for which some (but not all) of the 
virtues of a dynamical symmetry are retained. 
The essential idea is to relax the stringent conditions
of {\em complete} solvability
so that the properties (i)--(iii)
are only partially satisfied.
It is then possible to identify several types of 
partial dynamical symmetries. 
PDS of type~I corresponds to a situation where 
{\it some} of the states have {\it all} the dynamical symmetry. 
In this case, properties (i)-(iii) are fulfilled exactly, but 
by only a subset of states. 
PDS of type~II corresponds to a situation for which 
{\it all} the states preserve {\it part} of the dynamical symmetry. 
In this case, there are no analytic solutions,
yet selected quantum numbers (of the conserved symmetries) are retained. 
PDS of type~III has a hybrid character, for which {\it some} of the 
states preserve {\it part} of the dynamical symmetry. 
\begin{center}
\begin{table}[t]
\caption{\label{TabIBMcas}
\small
Generators, Casimir operators, $\hat{C}_{k}(G)$, of order 
$k=1,2,3$ and their eigenvalues for algebras~$G$ in the IBM. 
Here 
$\hat{n}_{s} = s^{\dag}s$, 
$\hat{n}_d = \sqrt{5}\,U^{(0)}$, $\hat{N}= \hat{n}_s + \hat{n}_d$, 
$\hat{L}_{m} = \sqrt{10}\,U^{(1)}_{m}$,  
$\hat{Q}_{m} = \Pi^{(2)}_{m} -{\textstyle\frac{\sqrt{7}}{2}}\,U^{(2)}_{m}$, 
$\Pi^{(2)}_{m} = d^{\dag}_{m}s + s^{\dag}\tilde{d}_{m}$, 
$\bar{\Pi}^{(2)}_{m} = i(d^{\dag}_{m}s - s^{\dag}\tilde{d}_{m})$, 
$U^{(\ell)}_{m} = (d^{\dag}\,\tilde{d})^{(\ell)}_{m}$, where 
$\tilde{d}_{m} = (-1)^{m}d_{-m}$.} 
\vspace{1mm}
\centering
\begin{tabular}{llll}
\br
& & &\\[-3mm]
Algebra & Generators & Casimir operator $\hat{C}_{k}(G)$ & 
Eigenvalues $\langle\hat{C}_{k}(G)\rangle$\\[4pt]
& & & \\[-3mm]
\mr
& & & \\[-2mm]
{\rm O(3)} & $U^{(1)}$ & $\hat{L}\cdot \hat{L}$ & L(L+1) \\[2pt]
{\rm O(5)} & $U^{(1)},U^{(3)}$ & 
$2(U^{(1)}\cdot U^{(1)} +U^{(3)}\cdot U^{(3)})$ & $\tau(\tau+3)$ \\[2pt]
{\rm O(6)} & $U^{(1)},U^{(3)},\Pi^{(2)}$ & 
$\hat{C}_{2}({\rm O(5)}) 
+ \Pi^{(2)}\cdot\Pi^{(2)}$ & $\Sigma(\Sigma+4)$ \\[2pt]
{\rm SU(3)} & $U^{(1)},\hat{Q}$ &
$2\hat{Q}\cdot \hat{Q} + {\textstyle\frac{3}{4}}\hat{L}\cdot \hat{L}$ &
$\lambda^2 +(\lambda+\mu)(\mu+3)$\\[2pt]
& &
$-4\,\sqrt{7}\hat{Q}\cdot (\hat{Q}\times \hat{Q}) ^{(2)}
-{\textstyle\frac{9}{2}\sqrt{3}}\hat{Q}\cdot 
(\hat{L}\times \hat{L})^{(2)}$  & 
$(\lambda -\mu)(2\lambda+\mu+3)$\\
& & & $\times(\lambda+2\mu+3)$\\[2pt]
{\rm U(5)} & $U^{(\ell)}$ $\ell=0,...\,,4$ & 
$\hat{n}_d,\,\hat{n}_d(\hat{n}_d+4)$ & $n_d,\, n_d(n_d+4)$ \\[2pt]
{\rm U(6)} & $U^{(\ell)}$ $\ell=0,...\,,4$ &
$\hat{N},\,\hat{N}(\hat{N}+5)$ & $N,\, N(N+5)$ \\[2pt]
      & $\Pi^{(2)},\, \bar{\Pi}^{(2)},\, \hat{n}_s$ & & \\[2pt]
& & &\\[-3mm]
\br
\end{tabular}
\label{Tab1}
\end{table}
\end{center}

In what follows we discuss algorithms for constructing Hamiltonians 
with partial dynamical symmetries and demonstrate 
their relevance to quantum systems. For that purpose, we 
employ the interacting boson model (IBM)~\cite{ibm}, 
widely used in the description of low-lying quadrupole collective states 
in nuclei in terms of $N$ interacting monopole $(s)$ and
quadrupole $(d)$ bosons representing valence nucleon pairs.
The bilinear combinations 
$\{s^{\dag}s,\,s^{\dag}d_{m},\, d^{\dag}_{m}s,\, 
d^{\dag}_{m}d_{m '}\}$ span a U(6) algebra, which 
serves as the spectrum generating algebra. 
The IBM Hamiltonian is expanded in terms of these generators 
and consists of Hermitian, rotational-scalar interactions 
which conserve the total number of $s$- and $d$- bosons, 
$\hat N = \hat{n}_s + \hat{n}_d = 
s^{\dagger}s + \sum_{m}d^{\dagger}_{m}d_{m}$. 
Three dynamical symmetry limits occur 
in the model with leading subalgebras U(5), SU(3), and O(6),
corresponding to typical collective spectra observed in nuclei,
vibrational, rotational, and $\gamma$-unstable, respectively.
Relevant information on these algebras is collected in Table~\ref{Tab1}. 
A geometric visualization of the model is obtained by 
an energy surface
\ba
E_{N}(\beta,\gamma) &=& 
\langle \beta,\gamma; N\vert \hat{H} \vert \beta,\gamma ; N\rangle ~, 
\label{enesurf}
\ea
defined by the expectation value of the Hamiltonian in the coherent 
(intrinsic) state~\cite{gino80,diep80}
\bsub
\ba
\vert\beta,\gamma ; N \rangle &=&
(N!)^{-1/2}(b^{\dagger}_{c})^N\,\vert 0\,\rangle ~,\\
b^{\dagger}_{c} &=& (1+\beta^2)^{-1/2}[\beta\cos\gamma 
d^{\dagger}_{0} + \beta\sin{\gamma} 
( d^{\dagger}_{2} + d^{\dagger}_{-2})/\sqrt{2} + s^{\dagger}] ~. 
\ea
\label{condgen}
\esub
Here $(\beta,\gamma)$ are
quadrupole shape parameters whose values, $(\beta_0,\gamma_0)$, 
at the global minimum of $E_{N}(\beta,\gamma)$ define the equilibrium 
shape for a given Hamiltonian. 
The shape can be spherical $(\beta =0)$ or 
deformed $(\beta >0)$ with $\gamma =0$ (prolate), $\gamma =\pi/3$ (oblate), 
$0 < \gamma < \pi/3$ (triaxial), or $\gamma$-independent. 
The equilibrium deformations associated with the 
dynamical symmetry limits are 
$\beta_0=0$ for U(5), $(\beta_0=\sqrt{2},\gamma_0=0)$ for SU(3) and 
$(\beta_0=1,\gamma_0\,{\rm arbitrary})$ for O(6). 

\section{Construction of Hamiltonians with partial dynamical symmetries}
\label{sec:PDSalgorithm}

PDS of type I corresponds to a situation for which the defining 
properties of a dynamical symmetry (DS), namely, solvability, 
good quantum numbers, and symmetry-dictated structure are fulfilled exactly, 
but by only a subset of states. 
An algorithm for constructing Hamiltonians with PDS 
has been developed in~\cite{AL92} and further elaborated 
in~\cite{RamLevVan09}. The analysis starts from the chain of nested algebras
\begin{equation}
\begin{array}{ccccccc}
G_{\rm dyn}&\supset&G&\supset&\cdots&\supset&G_{\rm sym}\\
\downarrow&&\downarrow&&&&\downarrow\\[0mm]
[h]&&\langle\Sigma\rangle&&&&\Lambda
\end{array}
\label{chain}
\end{equation}
where, below each algebra,
its associated labels of irreps are given. Eq.~(\ref{chain}) implies 
that $G_{\rm dyn}$ is the dynamical (spectrum generating) 
algebra of the 
system such that operators of all physical observables 
can be written in terms of its generators; 
a single irrep of $G_{\rm dyn}$
contains all states of relevance in the problem.
In contrast, $G_{\rm sym}$ is the symmetry algebra
and a single of its irreps contains states that are degenerate in energy. 
Assuming, for simplicity,
that particle number is conserved, then 
all states, and hence the representation $[h]$,
can then be assigned a definite particle number~$N$.
For $N$ identical particles the representation 
$[h]$ of the dynamical algebra 
$G_{\rm dyn}$ 
is either symmetric $[N]$ (bosons)
or antisymmetric $[1^N]$ (fermions)
and will be denoted, in both cases, as $[h_N]$. 
The occurrence of a DS of the type~(\ref{chain})
signifies that the Hamiltonian is written in terms of the Casimir 
operators of the algebras in the chain, 
\ba
\hat{H}_{DS} = \sum_{G} a_{G}\,\hat{C}(G) ~, 
\label{hDS}
\ea
and its eigenstates can be labeled as
$|[h_N]\langle\Sigma\rangle\dots\Lambda\rangle$;
additional labels (indicated by $\dots$)
are suppressed in the following.
The eigenvalues of the Casimir operators in these basis states 
determine the eigenenergies $E_{DS}([h_N]\langle\Sigma\rangle\Lambda)$ 
of $\hat{H}_{DS}$. Likewise, operators can be classified
according to their tensor character under~(\ref{chain})
as $\hat T_{[h_n]\langle\sigma\rangle\lambda}$.

Of specific interest in the construction of a PDS
associated with the reduction~(\ref{chain}),
are the $n$-particle annihilation operators $\hat T$ 
which satisfy the property
\begin{equation}
\hat T_{[h_n]\langle\sigma\rangle\lambda}
|[h_N]\langle\Sigma_0\rangle\Lambda\rangle=0 ~,
\label{anni}
\end{equation}
for all possible values of $\Lambda$
contained in a given irrep~$\langle\Sigma_0\rangle$ of $G$. 
Equivalently, this condition can be phrased in terms of the action 
on a lowest weight (LW) state of the G-irrep $\langle\Sigma_0\rangle$,  
\ba
\hat T_{[h_n]\langle\sigma\rangle\lambda}
|LW;\, [h_N]\langle\Sigma_0\rangle\rangle=0 ~,
\label{LW}
\ea 
from which states of good $\Lambda$ can be obtained by projection. 
Any $n$-body, 
number-conserving normal-ordered interaction
written in terms of these annihilation operators 
and their Hermitian conjugates (which transform as the
corresponding conjugate irreps), 
\ba
\hat{H}' = \sum_{\alpha,\beta} 
A_{\alpha\beta}\, \hat{T}^{\dag}_{\alpha}\hat{T}_{\beta} ~, 
\label{PS}
\ea
has a partial G-symmetry. This comes about since for 
arbitrary coefficients, $A_{\alpha\beta}$, $\hat{H}'$ 
is not a G-scalar, hence most of its eigenstates will be a 
mixture of irreps of G, yet relation~(\ref{anni}) ensures that a subset of 
its eigenstates $\vert [h_N]\langle\Sigma_0\rangle\Lambda\rangle$, 
are solvable and have good quantum numbers under the chain~(\ref{chain}). 
An Hamiltonian with partial dynamical symmetry is obtained by adding 
to $\hat{H}'$ the dynamical symmetry Hamiltonian, 
$\hat{H}_{DS}$~(\ref{hDS}), still preserving the solvability
of states with $\langle\Sigma\rangle=\langle\Sigma_0\rangle$, 
\ba
\hat{H}_{PDS} &=& \hat{H}_{DS} + \hat{H}' ~.
\label{hPDS}
\ea 
If the operators $\hat T_{[h_n]\langle\sigma\rangle\lambda}$ 
span the entire irrep $\langle\sigma\rangle$ of G, 
then the annihilation condition~(\ref{anni}) is satisfied
for all $\Lambda$-states in $\langle\Sigma_0\rangle$, 
if none of the $G$ irreps $\langle\Sigma\rangle$
contained in the $G_{\rm dyn}$ irrep $[h_{N-n}]$
belongs to the $G$ Kronecker product
$\langle\sigma\rangle\times\langle\Sigma_0\rangle$. 
So the problem of finding interactions
that preserve solvability
for part of the states~(\ref{chain})
is reduced to carrying out a Kronecker product. 
The arguments for choosing the special irrep 
$\langle\Sigma\rangle=\langle\Sigma_0\rangle$ in Eq.~(\ref{anni}), 
which contains the solvable states, are based on 
physical grounds. A~frequently encountered choice is the irrep which 
contains the ground state of the system. 
The above algorithm is applicable to any semisimple group. 

PDS of type II corresponds to a situation for which {\it all} the
states of the system preserve {\it part} of the dynamical symmetry, 
$G_0 \supset G_1 \supset G_2 \supset \ldots \supset  G_{n}$. 
In this case, there are no analytic solutions,
yet selected quantum numbers (of the conserved symmetries) are retained.
This occurs, for example, 
when the Hamiltonian contains interaction
terms from two different chains with
a common symmetry subalgebra~\cite{Lev86}, {\it e.g.}, 
\ba
G_0 \supset 
\left \{
\begin{array}{c}
G_1 \\
G_{1}'
\end{array}
\right \}\supset 
G_2 \supset \ldots \supset G_n ~.
\label{G0chains}
\ea
If $G_{1}$ and $G_{1}'$ are incompatible, {\it i.e.}, do not commute, 
then their irreps are mixed in the eigenstates of the Hamiltonian. 
On the other hand, since $G_2$ and its subalgebras are common to both 
chains, then the labels of their irreps remain as good quantum numbers.  

An alternative situation where PDS of type II occurs is when the 
Hamiltonian preserves only some of the symmetries $G_i$ in the 
DS chain and only their irreps are unmixed. 
A~systematic procedure for identifying interactions with such property 
was proposed in~\cite{isa99}. 
Let $G_1\supset G_2\supset G_3$ be a set of nested algebras which 
may occur anywhere in the chain, 
in-between the spectrum generating algebra $G_0$ and the invariant 
symmetry algebra $G_n$. The procedure is based on 
writing the Hamiltonian in terms of generators, $g_i$, of $G_1$, 
which do not belong to its subalgebra $G_2$. 
By construction, such Hamiltonian preserves the 
$G_1$ symmetry but, in general, not the $G_2$ symmetry, and hence will 
have the $G_1$ labels as good quantum numbers but will mix different 
irreps of $G_2$. The Hamiltonians can still conserve the $G_3$ labels 
{\it e.g.}, by choosing it to be a scalar of $G_3$. 
The procedure 
involves the identification of the tensor character under $G_2$ and $G_3$ 
of the operators $g_i$ and their products, $g_i g_{j}\ldots g_{k}$. 
The Hamiltonians obtained in this manner
belong to the integrity basis of $G_3$-scalar operators in 
the enveloping algebra of $G_1$ and, hence, their 
existence is correlated with their order. 

PDS of type III
combines properties of both PDS of type I and II. Such a generalized 
PDS~\cite{levisa02} has a hybrid character, for which {\it part} of the 
states of the system under study preserve {\it part} 
of the dynamical symmetry. 
In relation to the dynamical symmetry chain of Eq.~(\ref{chain}), 
with associated basis, $\vert [h_N]\langle\Sigma\rangle\Lambda\rangle$, 
this can be accomplished by relaxing the condition of Eq.~(\ref{anni}), 
so that it holds only for {\it selected} 
states $\Lambda$ contained in a given irrep $\langle\Sigma_0\rangle$ 
of $G$ and/or selected (combinations of) components $\lambda$ of the tensor 
$\hat T_{[h_n]\langle\sigma\rangle\lambda}$. Under such circumstances, 
let $G'\neq G_{sym}$ be a subalgebra of $G$ in the aforementioned 
chain, $G\supset G'$. 
In general, the Hamiltonians, constructed from 
these tensors, in the manner shown in Eq.~(\ref{PS}), 
are not invariant under $G$ nor $G'$. 
Nevertheless, they do posses the subset of solvable states, 
$|[h_N]\langle\Sigma_0\rangle\Lambda\rangle$, 
with good $G$-symmetry $\langle\Sigma_0\rangle$ 
(which now span only part of the corresponding $G$-irrep), 
while other states are mixed. At the same time, the symmetry associated 
with the subalgebra $G'$, is broken in all states (including the solvable 
ones). Thus, part of the eigenstates preserve part of the symmetry. 
These are precisely the requirements of PDS of type III.

\section{SU(3) partial dynamical symmetry}

\label{subsec:su3PDStypeI}

The SU(3) DS chain of the IBM 
and related quantum numbers are given by~\cite{ibm}
\ba
\begin{array}{ccccc}
{\rm U}(6)&\supset&{\rm SU}(3)&\supset&{\rm O}(3)\\
\downarrow&&\downarrow&&\downarrow\\[0mm]
[N]&&\left (\lambda,\mu\right )& K & L
\end{array} ~.
\label{chainsu3}
\ea 
For a given U(6) irrep $[N]$, the allowed SU(3) irreps are 
$(\lambda,\mu)=(2N-4k-6m,2k)$ 
with $k,m$ non-negative integers, such that, $\lambda,\mu\geq 0$. 
The multiplicity label $K$ 
is needed for complete classification and corresponds geometrically to the
projection of the angular momentum on the symmetry axis. 
The values of~$L$ contained in a given SU(3) irrep $(\lambda,\mu)$, 
are obtained from the known $SU(3)\supset O(3)$ reduction. 
The states $\vert [N](\lambda,\mu)KL\rangle$ 
form the (non-orthogonal) Elliott basis~\cite{Elliott58} 
and the Vergados basis 
$\vert [N](\lambda,\mu)\tilde{\chi}L\rangle$~\cite{ibm} is obtained 
from it by a standard orthogonalization procedure. 
The two bases coincide in the large-N limit and both are eigenstates 
of a Hamiltonian with SU(3) DS. The latter, 
for one- and two-body interactions, can be transcribed in the form 
\ba
\hat{H}_{DS} &=& h_{2}\left [-\hat C_{2}(\rm SU(3)) 
+ 2\hat N (2\hat N+3)\right ]
+ C\, \hat C_{2}(\rm O(3)) ~,
\label{hDSsu3}
\ea
where $\hat{C}_{2}(G)$ is   
the quadratic Casimir operator of $G$, as defined in Table~\ref{Tab1}. 
The spectrum of $\hat{H}_{DS}$ is completely solvable with eigenenergies
\ba
E_{\rm DS} &=& 
h_{2}\,6\left [2N(k+2m) - k(2k-1)-3m(2m-1) -6km\right ]
+ CL(L+1) ~,
\label{eDSsu3}
\ea
and $(\lambda,\mu)=(2N-4k-6m,2k)$. The spectrum 
resembles that of an axially-deformed rotovibrator and the corresponding  
eigenstates are arranged in SU(3) multiplets. 
In a given SU(3) irrep $(\lambda,\mu)$, each $K$-value is associated 
with a rotational band and states 
with the same L, in different $K$-bands, are degenerate. 
The lowest SU(3) irrep is $(2N,0)$, which describes the ground band 
$g(K=0)$ of a prolate deformed nucleus. 
The first excited SU(3) irrep
$(2N-4,2)$ contains both the $\beta(K=0)$ and $\gamma(K=2)$ bands. 
States in these bands with the same angular momentum are degenerate.
This $\beta$-$\gamma$ degeneracy is a characteristic feature of the SU(3) 
limit of the IBM which, however, is not commonly observed. 
In most deformed nuclei the $\beta$ band lies above the $\gamma$ band. 
In the IBM framework, with at most two-body interactions, one
is therefore compelled to break SU(3) 
in order to conform with the experimental data. 

The construction of Hamiltonians with SU(3)-PDS of type I 
is based on identification 
of $n$-boson operators which annihilate all states in a given 
SU(3) irrep $(\lambda,\mu)$, 
chosen here to be the ground band irrep $(2N,0)$. 
For that purpose, we consider the following 
two-boson SU(3) tensors, 
$B^{\dagger}_{[n](\lambda,\mu)\tilde{\chi};\ell m}$, with $n=2$, 
$(\lambda,\mu)=(0,2)$ and 
angular momentum $\ell =0,\,2$
\bsub
\ba
B^{\dagger}_{[2](0,2)0;00} &\propto& 
P^{\dagger}_{0} = d^{\dagger}\cdot d^{\dagger} - 2(s^{\dagger})^2 ~,\\
B^{\dagger}_{[2](0,2)0;2m} &\propto& 
P^{\dagger}_{2m} = 2d^{\dagger}_{m}s^{\dagger} + 
\sqrt{7}\, (d^{\dagger}\,d^{\dagger})^{(2)}_{m} ~.
\ea
\label{PL}
\esub 
The corresponding Hermitian conjugate boson-pair annihilation operators,  
$P_0$ and $P_{2m}$, transform 
as $(2,0)$ under SU(3), and satisfy
\ba
P_{0}\,\vert [N](2N,0)K=0, L\rangle &=& 0~,
\nonumber\\
P_{2m}\,\vert [N](2N,0)K=0, L\rangle &=& 0~, 
\;\;\;\;\;
L=0,2,4,\ldots, 2N~.
\label{P0P2}
\ea
Equivalently, these operators satisfy
\ba
P_{0}\vert\beta=\sqrt{2},\gamma=0 ; N \rangle &=& 0 ~,
\nonumber\\
P_{2m}\vert\beta=\sqrt{2},\gamma=0 ; N \rangle &=& 0 ~,
\label{PLcond}
\ea
where $\vert\beta=\sqrt{2},\gamma=0 ; N \rangle$, 
is the condensate of Eq.~(\ref{condgen}) with the SU(3) equilibrium 
deformations. It is the lowest-weight state in the SU(3) irrep 
$(\lambda,\mu)=(2N,0)$ and serves as an intrinsic state for the 
SU(3) ground band.
The rotational members of the band $\vert [N](2N,0)K=0, L\rangle$, 
Eq.~(\ref{P0P2}), are obtained from it by O(3) projection, and 
span the entire SU(3) irrep $(\lambda,\mu)=(2N,0)$. 
The relations in Eqs.~(\ref{P0P2})-(\ref{PLcond}) follow from the 
fact that the action of the operators $P_{\ell m}$ leads to a state with 
$N-2$ bosons in the U(6) irrep $[N-2]$, 
which does not contain the SU(3) irreps obtained from the product 
$(2,0)\times (2N,0)= (2N+2,0)\oplus (2N,1)\oplus (2N-2,2)$. 
In addition,  $P_0$ satisfies 
\ba
P_{0}\,\vert [N](2N-4k,2k)K=2k, L\rangle = 0 ~,\;\;\;\;
L=K,K+1,\ldots, (2N-2k)~.
\label{P0}
\ea
For $k> 0$ the indicated $L$-states 
span only part of the SU(3) irreps 
$(\lambda,\mu)=(2N-4k,2k)$ and form the rotational members of excited 
$\gamma^{k}(K=2k)$ bands. This result follows from the fact that $P_{0}$ 
annihilates the intrinsic states of these bands, 
$\vert\gamma^k(K=2k)\rangle\propto 
(P^{\dag}_{2,2})^{k}\vert\beta=\sqrt{2},\gamma=0 ; N-2k \rangle$.

Following the general algorithm, a two-body Hamiltonian with partial 
SU(3) symmetry can now be constructed as in Eq.~(\ref{PS}), 
$\hat{H}' = h_{0} P^{\dagger}_{0}P_{0} 
+ h_{2} P^{\dagger}_{2}\cdot \tilde{P}_{2}$, 
where $\tilde P_{2m} = (-)^{m}P_{2,-m}$. 
For $h_{2}=h_{0}$, this Hamiltonian is an SU(3) scalar, 
while for $h_0=-5h_2$, it transforms as a $(2,2)$ SU(3) tensor component.
The scalar part is related to the 
quadratic Casimir operator of SU(3) 
\ba
\hat{\theta}_2 \equiv
P^{\dagger}_{0}P_{0} + P^{\dagger}_{2}\cdot \tilde{P}_{2} = 
-\hat C_{2}(\rm SU(3)) + 2\hat N (2\hat N+3) ~,
\label{theta}
\ea 
and is simply the first term in $\hat{H}_{DS}$, Eq.~(\ref{hDSsu3}). 
In accord with Eq.~(\ref{hPDS}), 
the two-body SU(3)-PDS Hamiltonian is thus given by
\ba
\hat{H}_{PDS} &=& 
\hat{H}_{DS} + \eta\, P^{\dagger}_{0}P_{0} ~.
\label{hPDSsu3}
\ea
The $P^{\dag}_{0}P_0$ term is not diagonal in the SU(3) chain, however,  
Eqs.~(\ref{P0P2})-(\ref{P0}) ensure that 
$\hat{H}_{PDS}$ retains selected solvable states with good 
SU(3) symmetry. Specifically, the solvable states are members of the 
ground $g(K=0)$ 
\bsub
\ba
&&\vert N,(2N,0)K=0,L\rangle \;\;\;\; L=0,2,4,\ldots, 2N\\ 
&& E_{PDS}=CL(L+1)
\ea
\label{2N0}
\esub
and $\gamma^{k}(K=2k)$ bands
\bsub
\ba
&&\vert N,(2N-4k,2k)K=2k,L\rangle
\;\;\;\; 
L=K,K+1,K+2,\ldots, (2N-2k)\\
&& E_{PDS} = 
h_{2}\,6k(N-2)(2N -2k+1) + CL(L+1) \qquad\quad k>0~.
\ea
\label{2N4k2k}
\esub
The remaining eigenstates of $\hat{H}_{PDS}$ do 
not preserve SU(3) and, therefore, get mixed. 
\begin{figure}[t]
\begin{center}
\includegraphics[height=6cm]{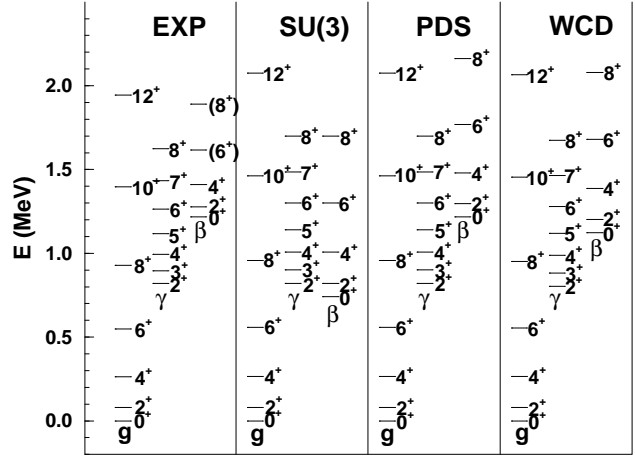}
\caption{
\small
Spectra of $^{168}$Er ($N=16$). Experimental energies
(EXP) are compared with IBM calculations in an exact SU(3) dynamical 
symmetry [SU(3)], in a broken SU(3) symmetry (WCD) 
and in a partial dynamical SU(3) symmetry (PDS). 
The latter employs the Hamiltonian of Eq.~(\ref{hPDSsu3}), 
with $h_2=4,\,\eta=4,\,C=13$ keV~\cite{lev96}. 
\label{fig1er168}}
\end{center}
\end{figure}
\begin{center}
\begin{table}[bh]
\caption{\label{Tab3Er168}
\small
B(E2) branching ratios from states in the $\gamma$ band in
$^{168}$Er. The column EXP lists the experimental ratios, 
PDS is the SU(3) partial dynamical symmetry 
calculation and WCD is a broken SU(3) calculation~\cite{lev96}.}
\vspace{1mm}
\centering
\begin{tabular}{lcccc|rlcccc}
\br
& & & & & & & & & &\\[-3mm]
$L^{\pi}_{i}$ & $L^{\pi}_{f}$ &  EXP &  PDS &  WCD &    &
$L^{\pi}_{i}$ & $L^{\pi}_{f}$ &  EXP &  PDS &  WCD \\
& & & & & & & & & &\\[-3mm]
\mr
& & & & & & & & & &\\[-2mm]
$2^{+}_{\gamma}$ & $0^{+}_{g}$      & $54.0$   &  $64.27$ &  $66.0$  &    &
$6^{+}_{\gamma}$ & $4^{+}_{g}$      &   $0.44$ &   $0.89$ &   $0.97$ \\[2pt]
                 & $2^{+}_{g}$      & $100.0$  & $100.0$  & $100.0$  &    &
                 & $6^{+}_{g}$      &   $3.8$  &   $4.38$ &   $4.3$  \\[2pt]
                 & $4^{+}_{g}$      &   $6.8$  &   $6.26$ &   $6.0$  &    &
                 & $8^{+}_{g}$      &   $1.4$  &   $0.79$ &   $0.73$ \\[2pt]
$3^{+}_{\gamma}$ & $2^{+}_{g}$      &   $2.6$  &   $2.70$ &   $2.7$  &    &
                 & $4^{+}_{\gamma}$ & $100.0$  & $100.0$  & $100.0$  \\[2pt]
                 & $4^{+}_{g}$      &   $1.7$  &   $1.33$ &   $1.3$  &    &
                 & $5^{+}_{\gamma}$ &  $69.0$  &  $58.61$ &  $59.0$  \\[2pt]
                 & $2^{+}_{\gamma}$ & $100.0$  & $100.0$  & $100.0$  &    &
$7^{+}_{\gamma}$ & $6^{+}_{g}$      &   $0.74$ &   $2.62$ &   $2.7$  \\[2pt]
$4^{+}_{\gamma}$ & $2^{+}_{g}$      &   $1.6$  &   $2.39$ &   $2.5$  &    &
                 & $5^{+}_{\gamma}$ & $100.0$  & $100.0$  & $100.0$  \\[2pt]
                 & $4^{+}_{g}$      &   $8.1$  &   $8.52$ &   $8.3$  &    &
                 & $6^{+}_{\gamma}$ &  $59.0$  &  $39.22$ &  $39.0$  \\[2pt]
                 & $6^{+}_{g}$      &   $1.1$  &   $1.07$ &   $1.0$  &    &
$8^{+}_{\gamma}$ & $6^{+}_{g}$      &   $1.8$  &   $0.59$ &   $0.67$ \\[2pt]
                 & $2^{+}_{\gamma}$ & $100.0$  & $100.0$  & $100.0$  &    &
                 & $8^{+}_{g}$      &   $5.1$  &   $3.57$ &   $3.5$  \\[2pt]
$5^{+}_{\gamma}$ & $4^{+}_{g}$      &   $2.91$ &   $4.15$ &   $4.3$  &    &
                 & $6^{+}_{\gamma}$ & $100.0$  & $100.0$  & $100.0$  \\[2pt]
                 & $6^{+}_{g}$      &   $3.6$  &   $3.31$ &   $3.1$  &    &
                 & $7^{+}_{\gamma}$ & $135.0$  &  $28.64$ &  $29.0$  \\[2pt]
                 & $3^{+}_{\gamma}$ & $100.0$  & $100.0$  & $100.0$  &    &
                 &                  &          &          &          \\[2pt]
                 & $4^{+}_{\gamma}$ & $122.0$  &  $98.22$ &  $98.5$  &    &
                 &                  &          &          &          \\[4pt]
\br
\end{tabular}
\label{Tabbe2su3}
\end{table}
\end{center}

The empirical spectrum of $^{168}$Er is shown in 
Fig.~\ref{fig1er168} and compared with SU(3)-DS, SU(3)-PDS and broken 
SU(3) calculations~\cite{lev96}. The SU(3)-PDS spectrum shows an
improvement over the schematic, exact SU(3) dynamical symmetry 
description, since the $\beta$-$\gamma$ degeneracy is lifted. 
The quality of the calculated PDS spectrum is similar to that obtained
in the broken-SU(3) calculation, however, in the former 
the ground $g(K=0_1)$ and $\gamma(K=2_1)$ bands remain solvable 
with good SU(3) symmetry, $(\lambda,\mu)=(2N,0)$ and $(2N-4,2)$ respectively. 
At the same time, the excited $K=0^{+}_2$ band involves about $13\%$ 
SU(3) admixtures into the dominant $(2N-4,2)$ irrep. 
Since the wave functions of the solvable 
states~(\ref{2N0})-(\ref{2N4k2k}) are known, one can obtain 
{\it analytic} expressions for matrix elements of observables between them. 
For example, the most general one-body E(2) operator reads 
$T(E2) = \alpha \hat{Q} + \rho \Pi^{(2)}$, 
in the notation of Table~\ref{Tab1}. Since $\hat{Q}$ is an SU(3) generator, 
it cannot connect different SU(3) irreps, hence only $\Pi^{(2)}$, 
which is a (2,2) SU(3) tensor, contributes to $\gamma\to g$ transitions. 
Accordingly, the calculated B(E2) ratios for $\gamma\to g$ transitions 
involve ratios of known SU(3) isoscalar factors and lead to parameter-free 
predictions. The latter, as shown in Table~\ref{Tab3Er168}, are 
in excellent agreement with experiment, thus 
confirming the relevance of SU(3)-PDS to the spectroscopy 
of $^{168}$Er~\cite{lev96}.

The construction of SU(3)-PDS Hamiltonians with higher-order terms 
follows the general algorithm and is based on identification 
of $n$-boson operators which annihilate all states in the irrep $(2N,0)$. 
For $n=3$, we consider the following SU(3) tensors, 
$\hat{B}^{\dagger}_{[n](\lambda,\mu)\tilde{\chi}; \ell m}$, 
\bsub
\ba
&&\hat B^\dag_{[3](2,2)0;00} 
\propto 
\;\;W^{\dag}_{0} = 5P^{\dag}_{0}s^{\dag}- P^{\dag}_{2}\cdot d^{\dag}
\;\;\; , \;\;\;
\hat B^\dag_{[3](2,2)2;2m} 
\propto 
W^{\dag}_{2m} = P^{\dag}_{0}d^{\dag}_{m} + 2P^{\dag}_{2m}s^{\dag}
\;\;\; ,\;
\nonumber\\
&&\hat B^\dag_{[3](2,2)0;2m} 
\propto 
V^{\dag}_{2m} = 6P^{\dag}_{0}d^{\dag}_{m} - P^{\dag}_{2m}s^{\dag}
\;\;\; , \;\;\;
\hat B^\dag_{[3](2,2)2;\ell m} 
\propto 
W^{\dag}_{\ell m} = (P^{\dag}_{2}d^{\dag})^{(\ell)}_{m}\;\;\; 
\ell=3,4 \qquad\quad
\label{WV}\\
&&\hat B^\dag_{[3](0,0)0;00} 
\propto 
\;\;\,\,\Lambda^{\dag} = P^{\dag}_{0}s^{\dag} + P^{\dag}_{2}\cdot d^{\dag} ~.
\label{Lam}
\ea
\label{B3}
\esub
The operators $W^{\dag}_{0},\,W^{\dag}_{2m},\,V^{\dag}_{2m},\,
W^{\dag}_{3m},\,W^{\dag}_{4m}$ in Eq.~(\ref{WV}) span the 
irrep $(\lambda,\mu)=(2,2)$, 
while $\Lambda^{\dagger}$ of Eq.~(\ref{Lam}) transforms as 
$(\lambda,\mu)=(0,0)$. 
The latter SU(3)-scalar operator is related to the cubic and quadratic 
Casimir operators of SU(3), defined in Table~\ref{Tab1},
\ba
2\Lambda^{\dag}\Lambda
&=& \hat{C}_{3}(SU(3)) - 2\hat N (4\hat{N}+3)(2\hat{N}+3)
+3(2\hat{N}+3)\hat{\theta}_2 ~, 
\ea
where $\hat{\theta}_{2}$ is given in Eq.~(\ref{theta}). 
In the presence of two- and three-body terms, the dynamical-symmetry 
Hamiltonian and eigenenergies for states with 
$(\lambda,\mu)=(2N-4k-6m,2k)$, read
\bsub
\ba
\hat{H}_{DS} &=& 
h_1\,\Lambda^{\dag}\Lambda + h_2\,\hat{\theta}_2 + 
C\,\hat{L}\cdot\hat{L} ~,\\
E_{DS} &=&  
+h_1\, 54m\left [ \,N(2k+2m+1) - k(2k-1) - (2m-1)(2m+1) -6km \,\right ]
\nonumber\\
&&
+ h_2\,6\left [\,2N(k+2m) - k(2k-1)-3m(2m-1) -6km \,\right ]
+ C\, L(L+1) ~.\qquad
\label{EDSa}
\ea
\label{EDS}
\esub

The SU(3)-PDS Hamiltonian has the following SU(3)-tensor expansion
\ba
\hat{H}_{PDS} &=& 
\hat{H}_{DS} + \eta\, P^{\dagger}_{0}P_{0} + 
a_{1}\,W^{\dag}_{0}W_{0} + 
a_{2}\,\left (W^{\dag}_{0}\Lambda + \Lambda^{\dag}W_{0}\right )
+ a_{3}\,W^{\dag}_{2}\cdot\tilde{W}_{2}
\nonumber\\
&&
+ a_{4}\,V^{\dag}_{2}\cdot\tilde{V}_{2}
+ a_{5}\,\left (W^{\dag}_{2}\cdot\tilde{V}_{2}
+ V^{\dag}_{2}\cdot\tilde{W}_{2}\right )
+a_{6}\,W^{\dag}_{3}\cdot \tilde{W}_{3} 
+a_{7}\,W^{\dag}_{4}\cdot \tilde{W}_{4} ~.\qquad
\ea
The relations of Eq.~(\ref{P0P2}) ensure that the operators 
$W_{\ell m}$, $V_{2m}$ and $\Lambda$ of Eq.~(\ref{B3}) annihilate the 
states of the SU(3) ground band $g(K=0)$. 
The solvable eigenstates and eigenenergies 
of $\hat{H}_{PDS}$ are those shown in Eq.~(\ref{2N0}). 
In addition, the operator $\Lambda$ (\ref{Lam}) 
annihilates {\it all} states in the irreps $(2N-4k,2k)$
\ba
\Lambda\,\vert [N](2N-4k,2k)K; L\rangle &=& 0 ~.
\label{Lam0}
\ea
This property follows from the fact that the U(6) irrep $[N-3]$ 
does not contain SU(3) irreps obtained from the product 
$(0,0)\times (2N-4k,2k)$. 
Using Eqs.~(\ref{P0}) and (\ref{Lam0}), we can identify the following 
sub-class of SU(3)-PDS Hamiltonians 
\ba
\hat{H}_{PDS-1} &=& \hat{H}_{DS} 
+h_{3}\,P^{\dag}_{0}P_{0} + 
h_{4}\, P^{\dag}_{0}s^{\dag}sP_{0}
+ h_{5}\,\left ( \Lambda^{\dag}sP_{0} + 
P^{\dag}_{0}s^{\dag}\Lambda\right ) ~,
\label{hPDS-1}
\ea
with additional solvable states which are 
the members of the $g(K=0)$ and $\gamma^{k}(K=2k)$ bands 
listed in Eqs.~(\ref{2N0})-(\ref{2N4k2k}).

A second sub-class of SU(3)-PDS Hamiltonian corresponds to the choice
\ba
\hat{H}_{PDS-2} &=& \hat{H}_{DS}
+h_{6}\,W^{\dag}_{2}\cdot \tilde{W}_{2} 
+h_{7}\,W^{\dag}_{3}\cdot \tilde{W}_{3} ~.
\label{hPDS-2}
\ea
$\hat{H}_{PDS-2}$ has a solvable ground band $g(K=0)$, Eq.~(\ref{2N0}), 
and a solvable $\beta(K=0)$ band 
\bsub
\ba
&&\vert N,(2N-4,2)K=0,L\rangle
\;\;\;\; 
L=0,2,4,\ldots, (2N-4)\\
&& E_{PDS-2} = 
h_{2}\,6(N-2)(2N -1) + CL(L+1) ~.
\ea
\esub
This result follows from the fact that $W_{2m}$ and $W_{3m}$ 
annihilate the intrinsic state of this band, 
$\vert \beta(K=0)\rangle \propto (\sqrt{2}P^{\dag}_{0} - P^{\dag}_{2,0})
\vert\beta=\sqrt{2},\gamma=0 ; N-2 \rangle$, as well as 
the projected Elliott basis states 
\ba
W_{\ell m}\vert [N](2N-4,2),K=0,L\rangle &=& 0 \qquad \ell=2,3~.
\ea

Three-body terms allow an additional solvable symmetry-conserving operator 
\ba
\hat{\Omega} &=& -4\sqrt{3}\, \hat{Q}\cdot 
(\hat{L}\times \hat{L})^{(2)} ~.
\label{Omega}
\ea
This operator is constructed from SU(3) generators, hence is diagonal 
in $(\lambda,\mu)$. It breaks, however, the aforementioned $K$-degeneracy. 
A well defined procedure exists for obtaining the eigenstates 
of $\hat{\Omega}$ and corresponding 
eigenvalues $\langle \hat{\Omega} \rangle$~\cite{Meyer85,Berghe85}. 
For example, 
for the irreps $(\lambda,0)$ and $(\lambda,2)$ with $\lambda$ even, 
we have
\bsub
\ba
&&(\lambda,0)\;  K = 0,\,\;\;\;\;\; L=0,2,4,\ldots,\lambda:
\qquad\qquad\;\;\;
\langle \hat{\Omega}\rangle = (2\lambda +3)L(L+1)~,
\label{ev0}
\\
&&(\lambda,2)\;  K= 2,\,\;\;\;\;\; L=3,5,7,
\ldots,\lambda + 1,\lambda+2: 
\quad
\langle \hat{\Omega}\rangle = (2\lambda +5)[ L(L+1) - 12]~,\;\;\\
&&(\lambda,2)\; K = 0,\,\;\;\;\;\; L=0: 
\qquad\qquad\qquad\qquad\quad
 \langle \hat{\Omega}\rangle = 0~,\\
&&(\lambda,2)\; K = 0,2,\;\;\, L=2,4,6,\ldots,\lambda:
\nonumber\\
&& \qquad\qquad\langle \hat{\Omega}\rangle = (2\lambda +5)[ L(L+1) - 6]
\pm 6 \sqrt{(2\lambda+5)^2 + L(L+1)(L-1)(L+2)}~.
\qquad\;
\ea
\label{Omegev}
\esub
Several works have examined the influence of the symmetry-conserving 
operator $\hat{\Omega}$ (\ref{Omega}) on nuclear spectra,  
within the IBM~\cite{Berghe85,Bona88,Vant90} and the symplectic 
shell model~\cite{RDW84,DR85}. It is interesting to note that the 
operator $\hat{\Omega}$ can be expressed in terms of 
$\hat{H}_{PDS-1}$~(\ref{hPDS-1}) and 
$\hat{H}_{PDS-2}$~(\ref{hPDS-2}) as
\ba
\hat{\Omega}
&=& -2(2\hat{N}-1)\hat{\theta}_2 + 2 \Lambda^{\dag}\Lambda 
+ (4\hat{N}+3)\hat{L}\cdot \hat{L} 
\nonumber\\
&&\;
+6(\hat{N}-1)P^{\dag}_{0}P_{0} 
-2\left ( \Lambda^{\dag}sP_{0} + P^{\dag}_{0}s^{\dag}\Lambda\right )
+2\,W^{\dag}_{2}\cdot \tilde{W}_{2} 
+4\,W^{\dag}_{3}\cdot \tilde{W}_{3} ~. 
\label{Omegap}
\ea

\section{O(6) partial dynamical symmetry}
\label{subsec:o6PDS}

The O(6) DS chain of the IBM and related quantum numbers are given 
by~\cite{ibm}
\ba
\begin{array}{ccccccc}
{\rm U}(6)&\supset&{\rm O}(6)&\supset&{\rm O}(5)&
\supset&{\rm O}(3)\\
\downarrow&&\downarrow&&\downarrow&&\downarrow\\[0mm]
[N]&&\langle\Sigma\rangle&&(\tau)&n_\Delta& L
\end{array} ~,
\label{chaino6}
\ea
For a given U(6) irrep $[N]$, the allowed O(6) and O(5) irreps are 
$\Sigma=N,\,N-2,\dots 0$ or $1$, and  
$\tau=0,\,1,\,\ldots \Sigma$, respectively. 
The values of $L$ contained in the O(5) 
irrep $(\tau)$ are obtained from the known 
${\rm O(5)}\supset {\rm O(3)}$ reduction and $n_{\Delta}$ is 
a multiplicity label. 
The eigenstates $|[N]\langle\Sigma\rangle(\tau)n_\Delta L\rangle$
are obtained with a Hamiltonian
with O(6) DS which, for one- and two-body interactions, can be
transcribed in the form
\ba
\hat{H}_{\rm DS} &=& h_{0}\left [-\hat C_{2}({\rm O(6)}) 
+ \hat N (\hat{N} +4)\right ]
+ B\, \hat{C}_{2}({\rm O(5)}) + C\,\hat{C}_{2}({\rm O(3)}) ~. 
\label{hDSo6}
\ea
Here the quadratic Casimir operators, $\hat{C}_{2}(G)$, are defined 
in Table~\ref{Tab1}. 
The spectrum of $\hat{H}_{\rm DS}$ is completely solvable with 
eigenenergies
\ba
E_{\rm DS} &=& 
4h_{0}\,(N-v +2)v + B\,\tau(\tau+3) +\, C\,L(L+1) ~.
\label{eDSo6v}
\ea
The spectrum resembles that of a $\gamma$-unstable deformed rotovibrator, 
where states are arranged in bands with O(6) quantum number
$\Sigma=N-2v$, $(v=0,1,2,\ldots)$.
The ground band ($v=0$) corresponds to the O(6) irrep with $\Sigma=N$. 
The O(5) and O(3) terms in $\hat{H}_{\rm DS}$~(\ref{hDSo6}), 
govern the in-band rotational splitting. 
The lowest members in each band have quantum numbers 
$(\tau=0,\, L=0)$, $(\tau=1,\, L=2)$ and $(\tau=2,\, L=2,4)$. 

The construction of Hamiltonians with O(6)-PDS of type I 
is based on identification 
of $n$-boson operators which annihilate all states in a given 
O(6) irrep, $\langle\Sigma\rangle$, chosen here to be the ground band 
irrep $\langle\Sigma\rangle=\langle N\rangle$. 
For that purpose, a relevant operator to consider is
\ba
\hat B^\dag_{[2]\langle0\rangle(0)0;00} &\propto& 
P^{\dagger}_{0} = d^{\dagger}\cdot d^{\dagger} - (s^{\dagger})^2 ~.
\label{Pdag0o6}
\ea
The corresponding Hermitian conjugate boson-pair annihilation operator, 
$P_0$, transforms also as $\langle\Sigma\rangle=\langle 0\rangle$ 
under O(6) and satisfies 
\ba
P_{0}\,\vert [N]\langle N \rangle (\tau)n_{\Delta}L\rangle = 0 ~, 
\;\;\;\;\; \tau=0,1,2,\ldots,N~.
\label{P0o6}
\ea
Equivalently, this operator satisfies
\ba
P_{0}\vert\beta=1,\gamma ; N \rangle &=& 0 ~,
\label{P0cond}
\ea
where $\vert\beta=1,\gamma ; N \rangle$,
is the condensate of Eq.~(\ref{condgen})
with the O(6) equilibrium deformations. 
It is the lowest-weight state in the O(6) irrep 
$\langle\Sigma\rangle = \langle N\rangle$ 
and serves as an intrinsic state 
for the O(6) ground band. The rotational members of the band, 
$\vert [N]\langle N\rangle (\tau)n_{\Delta}L\rangle$, Eq.~(\ref{P0o6}), 
are obtained from it by O(5) projection, 
and span the entire O(6) irrep $\langle\Sigma\rangle=\langle N\rangle$.
The relations in Eqs.~(\ref{P0o6})-(\ref{P0cond}) follow from the 
fact that the action of the operator $P_{0}$ leads to a state with 
$N-2$ bosons in the U(6) irrep $[N-2]$, 
which does not contain the O(6) irrep $\langle N\rangle$  obtained from the 
product of $\langle 0\rangle\times \langle N\rangle$.

Since both $P^{\dag}_{0}$ and $P_0$~(\ref{Pdag0o6}) 
are O(6) scalars, they give rise to the following interaction
\ba
P^{\dag}_{0}P_{0} &=& 
-\hat C_{{\rm O(6)}} 
+ \hat N (\hat{N} +4) ~,
\label{HPSo6}
\ea
which is simply the O(6) term in 
$\hat{H}_{\rm DS}$, Eq.~(\ref{hDSo6}), 
with an exact O(6) symmetry. Thus, in this case, 
unlike the situation encountered with SU(3)-PDS, 
the algorithm does not yield an O(6)-PDS of type I with two-body 
interactions. In the IBM framework, an Hamiltonian with a genuine 
O(6)-PDS of this class, requires higher-order terms. 

Focusing on three-body interactions with O(6)-PDS, 
one considers the following two three-boson O(6) tensors, 
$\hat B^\dag_{[n]\langle\sigma\rangle(\tau)n_{\Delta}\ell m}$, 
with $n=3$, $\sigma=1$ and $\ell=0,2$, 
\bsub
\ba
\hat B^\dag_{[3]\langle1\rangle(0)0;00}
&\propto& P^{\dag}_{0}s^\dag ~,\\
\hat B^\dag_{[3]\langle1\rangle(1)0;2m}
&\propto& P^{\dag}_{0}d^\dag_m ~.
\ea
\esub
\begin{figure*}[t]
\begin{center}
\leavevmode
\includegraphics[width=0.90\linewidth]{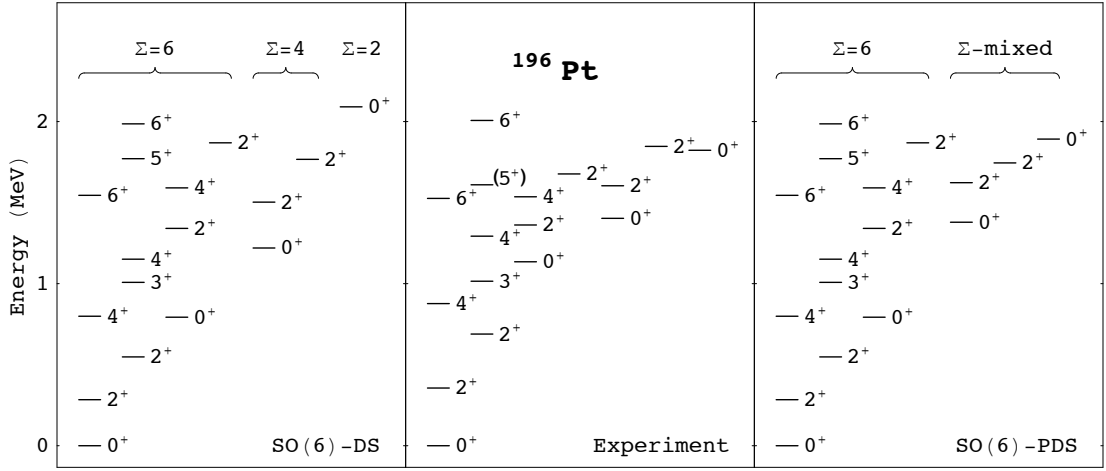}
\caption{
\small
Observed spectrum of $^{196}$Pt compared with the calculated spectra 
of $\hat H_{\rm DS}$~(\ref{hDSo6}), 
with O(6) dynamical symmetry (DS), 
and of $\hat H_{\rm PDS}$~(\ref{hPDSo63bod}) with 
O(6) partial dynamical symmetry (PDS). 
The parameters in $\hat H_{\rm DS}$ $(\hat H_{\rm PDS})$ are
$h_0=43.6\, (30.7)$, $B=44.0\, (44.0)$, $C=17.9\, (17.9)$, 
and $\eta=0\, (8.7)$ keV. The boson number is $N=6$ 
and $\Sigma$ is an O(6) label~\cite{RamLevVan09}.}
\label{fig2pt196}
\end{center}
\end{figure*}
\begin{center}
\begin{table}[b]
\caption{\label{Tabbe2pt196}
\small
Observed (EXP) and calculated B(E2) values 
(in $e^2{\rm b}^2$) for $^{196}$Pt. 
For both the exact (DS) and partial (PDS)
O(6) dynamical symmetry calculations, the E2 operator is
$T(E2) = e_{B}[ \, \Pi^{(2)} + \chi\,U^{(2)} \, ]$ 
with $e_{B}=0.151$ $e$b and $\chi=0.29$. Only the state $0^{+}_3$ 
has a mixed O(6) character~\cite{RamLevVan09}.} 
\vspace{1mm}
\centering
\begin{tabular}{clll|clll}
\br
& & & & & & &\\[-3mm] 
Transition& EXP & DS &PDS &
Transition& EXP & DS &PDS\\
& & & & & & &\\[-3mm] 
\mr
& & & & & & &\\[-2mm] 
$2^+_1\rightarrow0^+_1$& 0.274~(1)  &  0.274&  0.274 &
$2^+_3\rightarrow0^+_2$& 0.034~(34) &  0.119&  0.119\\[2pt]
$2^+_2\rightarrow2^+_1$& 0.368~(9)  &  0.358&  0.358 &
$2^+_3\rightarrow4^+_1$& 0.0009~(8) &  0.0004& 0.0004\\[2pt]
$2^+_2\rightarrow0^+_1$& 3.10$^{-8}$(3) & 0.0018& 0.0018 &
$2^+_3\rightarrow2^+_2$& 0.0018~(16)& 0.0013& 0.0013\\[2pt]
$4^+_1\rightarrow2^+_1$& 0.405~(6)  &  0.358&  0.358 &
$2^+_3\rightarrow0^+_1$& 0.00002~(2)& 0        & 0         \\[2pt]
$0^+_2\rightarrow2^+_2$& 0.121~(67) &  0.365&  0.365 &
$6^+_2\rightarrow6^+_1$& 0.108~(34) & 0.103& 0.103\\[2pt]
$0^+_2\rightarrow2^+_1$& 0.019~(10) &  0.003&  0.003 &
$6^+_2\rightarrow4^+_2$& 0.331~(88) &  0.221&  0.221\\[2pt]
$4^+_2\rightarrow4^+_1$& 0.115~(40) &  0.174&  0.174 &
$6^+_2\rightarrow4^+_1$& 0.0032~(9) & 0.0008& 0.0008\\[2pt]
$4^+_2\rightarrow2^+_2$& 0.196~(42) &  0.191&  0.191 &
$0^+_3\rightarrow2^+_2$&$<0.0028$   & 0.0037& 0.0028\\[2pt]
$4^+_2\rightarrow2^+_1$& 0.004~(1)  &  0.001&  0.001 &
$0^+_3\rightarrow2^+_1$&$<0.034$    & 0          & 0        \\[2pt]
$6^+_1\rightarrow4^+_1$& 0.493~(32) &  0.365&  0.365 &
                       &            &            &  \\[4pt]
\br
\end{tabular}
\end{table} 
\end{center}
The relation of Eq.~(\ref{P0o6}) ensures that $sP_{0}$ and $d_{m}P_0$ 
annihilate the states of the O(6) ground band. 
The only three-body interactions that are partially solvable in O(6)
are thus $P^{\dag}_{0}\hat n_s P_{0}$
and $P^{\dag}_{0}\hat n_d P_{0}$. 
Since the combination $P^{\dag}_{0}(\hat n_s+\hat n_d)P_{0}
= (\hat{N} -2)P^{\dag}_{0}P_{0}$
is completely solvable in O(6), we can transcribe the O(6)-PDS 
Hamiltonian in the form
\ba
\hat{H}_{\rm PDS}=\hat{H}_{\rm DS} + 
\eta\,P^{\dag}_{0}\hat{n}_s P_{0} ~,
\label{hPDSo63bod}
\ea
Here the dynamical symmetry Hamiltonian, $\hat{H}_{DS}$,  is that 
of Eq.~(\ref{hDSo6}), since no new terms 
are added to it at the level of three-body interactions.
The solvable states are members of the $\gamma$-unstable deformed 
ground band
\bsub
\ba
&& 
\vert [N]\langle N \rangle (\tau)n_{\Delta}L\rangle ~,
\;\;\;\;\; \tau=0,1,2,\ldots,N\\
&& E_{PDS} = B\tau(\tau+3) + CL(L+1) ~.
\label{ePDSo6I}
\ea
\esub

The experimental spectrum and E2 rates of $^{196}$Pt
are shown in Fig.~\ref{fig2pt196} and Table~\ref{Tabbe2pt196}.
The O(6)-DS limit is seen to provide a good description
for properties of states in the ground band $(\Sigma=N)$.
This observation was the basis of the claim~\cite{Cizewski78} 
that the O(6)-DS is manifested empirically in $^{196}$Pt.
However, the resulting fit to energies of excited bands is quite poor.
The $0^+_1$, $0^+_3$, and $0^+_4$ levels of $^{196}$Pt
at excitation energies 0, 1403, 1823 keV, respectively,
are identified as the bandhead states
of the ground $(v=0)$, first- $(v=1)$
and second- $(v=2)$ excited vibrational bands~\cite{Cizewski78}.
Their empirical anharmonicity,
defined by the ratio $R=E(v=2)/E(v=1)-2$,
is found to be $R=-0.70$.
In the O(6)-DS limit these bandhead states
have $\tau=L=0$ and $\Sigma=N,N-2,N-4$, respectively.
The anharmonicity $R=-2/(N+1)$,
as calculated from Eq.~(\ref{eDSo6v}), is fixed by $N$.
For $N=6$, which is the appropriate boson number for $^{196}$Pt,
the O(6)-DS value is $R=-0.29$,
which is in marked disagreement with the empirical value.
A detailed study of double-phonon excitations within the IBM,
has concluded that large anharmonicities can be incorporated
only by the inclusion of at least cubic terms in the
Hamiltonian~\cite{ramos00b}. 
In the IBM there are 17 possible three-body interactions~\cite{ibm}.
One is thus confronted with the need to select suitable higher-order terms
that can break the DS in excited bands but preserve it in the ground band. 
On the basis of the preceding discussion this can be accomplished by the 
O(6)-PDS Hamiltonian of Eq.~(\ref{hPDSo63bod}). 
The spectrum of $\hat{H}_{\rm PDS}$ is shown in Fig.~\ref{fig2pt196}.
The states belonging to the $\Sigma=N=6$ multiplet remain solvable
with energies (\ref{ePDSo6I}), which obey the same DS expression, 
Eq.~(\ref{eDSo6v}). States with $\Sigma < 6$ are generally admixed 
but agree better with the data than in the DS calculation. 
For example, the bandhead states of the first- (second-) excited bands 
have the O(6) decomposition 
$\Sigma=4$: $76.5\%\,(19.6\%)$, 
$\Sigma=2$: $16.1\%\,(18.4\%)$, 
and $\Sigma=0$: $7.4\%\,(62.0\%)$. 
Thus, although the ground band is pure, 
the excited bands exhibit strong O(6) breaking. 
The calculated O(6)-PDS anharmonicity for these bands is $R=-0.63$, 
much closer to the empirical value, $R=-0.70$. 
It should be emphasized that not only the energies 
but also the wave functions of the $\Sigma=N$ states remain unchanged
when the Hamiltonian is generalized from DS to PDS.
Consequently, the E2 rates for transitions among this class of states
are the same in the DS and PDS calculations. Thus, the additional 
three-body term in the Hamiltonian~(\ref{hPDSo63bod}), does not spoil 
the good O(6)-DS description for this segment of the spectrum. 
This is evident in Table~\ref{Tabbe2pt196}
where most of the E2 data concern transitions between $\Sigma=N=6$ states.

\section{Coexistence of partial dynamical symmetries}
\label{sec:PDSQPT}

The examples considered in previous sections involved Hamiltonians with 
a single PDS, describing stable structures, {\it e.g.}, well-deformed 
nuclei. Multiple partial dynamical symmetries can occur in 
systems undergoing quantum phase transitions (QPTs)~\cite{lev07}. 
The latter are structural changes induced by a variation of 
parameters in the Hamiltonian. 
Such ground-state phase transitions are a pervasive phenomenon 
observed in many branches of physics~\cite{carr10}, and are realized 
empirically in nuclei as transitions between different shapes. 
In the IBM, 
the dynamical symmetry limits correspond to possible phases of the system 
and the relevant Hamiltonians for studying shape-phase transitions 
involve terms with from different DS chains~\cite{diep80}. 
The nature of the phase transition is 
governed by the topology of the 
surface $E_{N}(\beta,\gamma)$, Eq.~(\ref{enesurf}), 
which serves as a Landau's potential 
with the equilibrium deformations as order parameters. 
The surface at the critical-point of a first-order transition 
is required to have two degenerate minima, corresponding to the two 
coexisting phases. Specifically, the first-order critical surface is 
\ba
E_{N}(\beta,\gamma=0) &=& 
2h_{2}N(N-1)(1+\beta^2)^{-2}\beta^2\left ( \beta-\beta_0\,\right )^2 ~,
\ea 
and has degenerate spherical and 
deformed minima at $\beta=0$ and $(\beta=\beta_0,\gamma=0)$, 
corresponding to spherical and axially-deformed shapes. 
A barrier of height $h = h_{2}N(N-1)(1-\sqrt{1+\beta_{0}^2})^{2}/2$ 
separates the two minima. 
Such a surface can be obtained from Eq.~(\ref{enesurf}) with 
the following critical-point Hamiltonian
\ba
\hat{H}_{cri}(\beta_0) &=& h_{2}\, 
P^{\dagger}_{2}(\beta_0)\cdot\tilde{P}_{2}(\beta_0) ~,
\;\;\;\;
P^{\dagger}_{2m}(\beta_0) = 
\beta_{0}\sqrt{2}d^{\dagger}_{m}s^{\dagger} + 
\sqrt{7}\, (d^{\dagger}\,d^{\dagger})^{(2)}_{m} ~.
\label{hcri1st}
\ea
$P_{2m}(\beta_0)$ annihilates the condensate of 
Eq.~(\ref{condgen}) with $(\beta=\beta_0,\gamma=0)$ as well as the states 
of good $L$, $\vert\, \beta_0; N, L \rangle$, projected from it
\bsub
\ba
P_{2m}(\beta_0)\,\vert\,\beta_0,\gamma=0 ; N \rangle &=& 0~,\\
P_{2m}(\beta_0)\,\vert\, \beta_0 ; N, L \rangle &=& 0~,
\;\;\;\;\;
L=0,2,4,\ldots, 2N~.
\label{P2b0}
\ea
\esub
Consequently, $\hat{H}_{cri}(\beta_0)$ has a solvable zero-energy 
prolate-deformed ground band, 
composed of the $L$-projected states mentioned above
\ba
\vert\beta_0;N,L\rangle \quad E=0\qquad\;\; L=0,2,4,\ldots, 2N~.
\qquad\quad
\label{deform}
\ea

The multipole form of $\hat{H}_{cri}(\beta_0)$ (\ref{hcri1st}) 
is given by
\ba
\hat{H}_{cri}(\beta_0) = 
h_{2}\left [2(\beta_{0}^2\hat{N} -2)\hat{n}_d 
+ 2(1-\beta_{0}^2)\hat{n}_{d}^2 
+ 2 \hat{C}_{2}({\rm O(5)}) - \hat{C}_{2}({\rm O(3)})
+ \sqrt{14}\beta_{0} \Pi^{(2)}\cdot U^{(2)}\right ],\,
\label{hcri1stmult}
\ea
where the various operators are defined in Table~\ref{Tab1}. 
The $\hat{n}_d,\,\hat{n}_{d}^2,\,\hat{C}_{2}(O(5))$ and 
$\hat{C}_{2}(O(3))$ terms in Eq.~(\ref{hcri1stmult}) 
belong to the dynamical symmetry Hamiltonian 
\begin{figure}[t]
\begin{minipage}{18pc}
\includegraphics[width=2.1in,angle=270,clip=]
{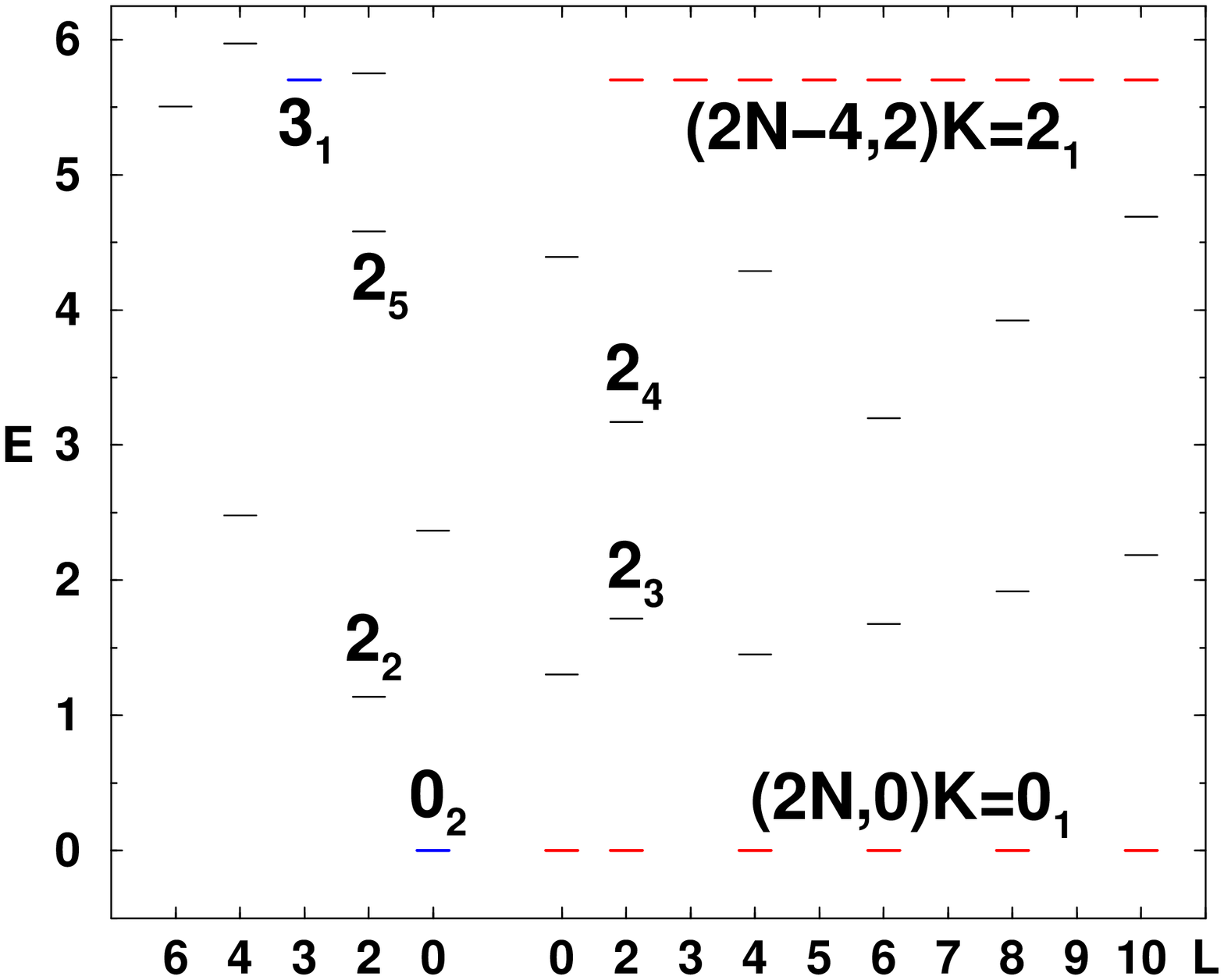}
\caption{\label{fig3}
\small
Spectrum of $\hat{H}_{cri}(\beta_0=\sqrt{2})$, 
Eq.~(\ref{hcri1st}), with $h_2=0.05$, $N=10$. 
$L(K=0_1)$ and $L(K=2_1)$ are the solvable SU(3) states 
of Eqs.~(\ref{2N0})-(\ref{2N4k2k}) with $k=0$ and $k=1$, respectively. 
$L=0_2,3_1$ are the solvable U(5) states of 
Eq.~(\ref{spher})~\cite{lev07}.}
\end{minipage}\hspace{0.7cm}%
\begin{minipage}{17pc}
\includegraphics[width=2.1in,angle=270,clip=]
{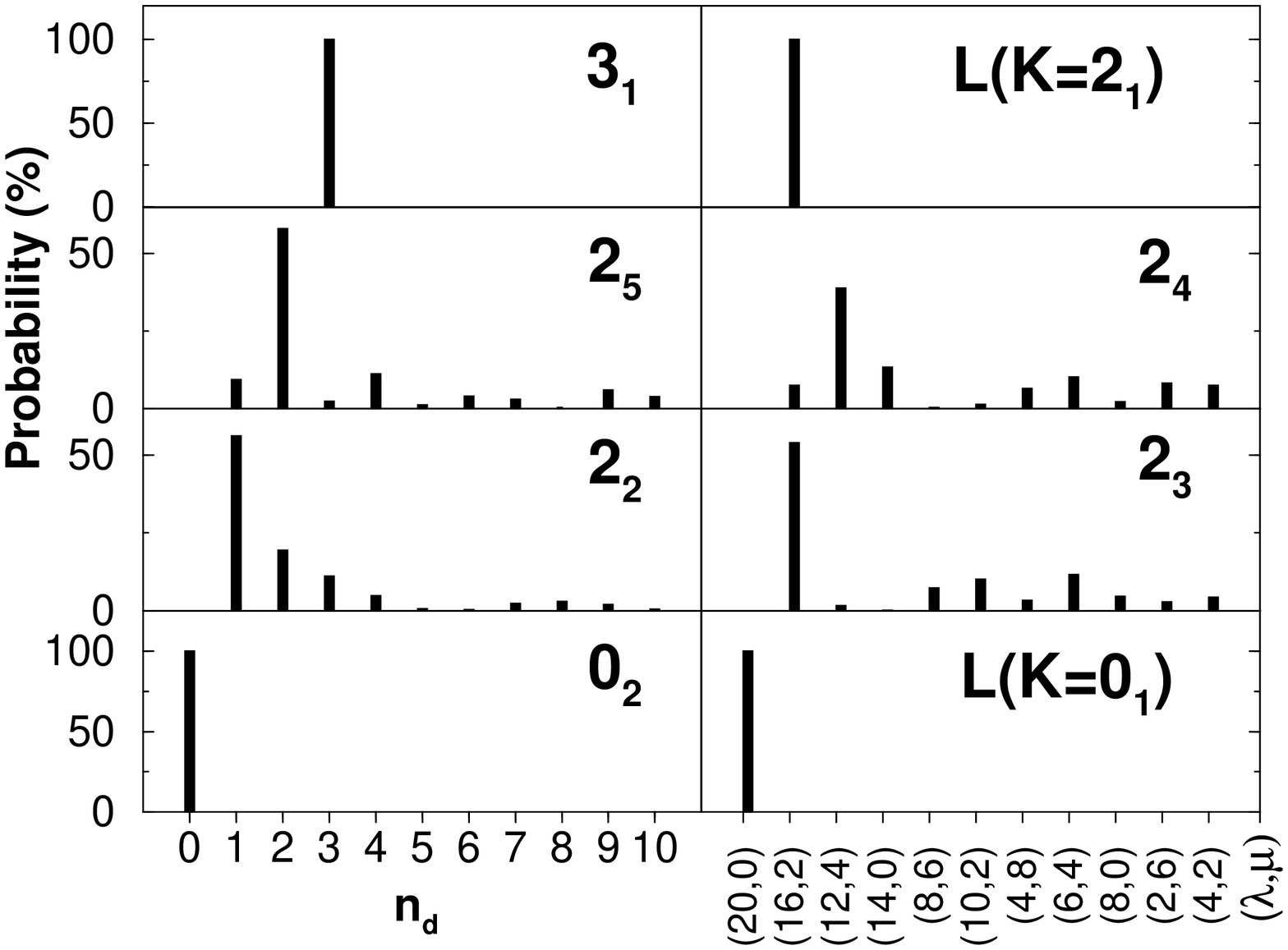}
\caption{\label{fig4}
\small
U(5) ($n_d$) and SU(3) $[(\lambda,\mu)]$ decomposition for 
selected spherical and deformed states in Fig.~\ref{fig3}~\cite{lev07}.}
\end{minipage} 
\end{figure} 
of the U(5) chain
\ba
\begin{array}{ccccccc}
{\rm U}(6)&\supset&{\rm U}(5)&\supset&{\rm O}(5)&
\supset&{\rm O}(3)\\
\downarrow&&\downarrow&&\downarrow&&\downarrow\\[0mm]
[N]&&\langle n_d \rangle&&(\tau)&n_\Delta& L
\end{array} ~,
\label{chainu5}
\ea 
and describe the dynamics of an anharmonic spherical vibrator. 
The $\Pi^{(2)}\cdot U^{(2)}$ term 
can connect states with $\Delta n_d = \pm 1$ 
and $\Delta\tau=\pm1,\pm 3$, 
hence breaks the U(5) DS.  
Nevertheless, $\hat{H}_{cri}(\beta_0)$ 
has selected solvable eigenstates with good U(5) symmetry, 
\bsub
\ba 
\vert N,n_d=\tau=L=0 \rangle  \;\; &&E = 0 ~,
\label{nd0b0}\\
\vert N,n_d=\tau=L=3 \rangle \;\;
&&E = 6 h_2[\beta_{0}^2 (N-3) + 5]~,
\qquad\quad
\label{nd3b0}
\ea
\label{spher}
\esub 
and therefore, by construction, $\hat{H}_{cri}(\beta_0)$ 
exhibits U(5)-PDS of type I.

For $\beta_0=\sqrt{2}$, 
the critical Hamiltonian of Eq.~(\ref{hcri1st}) 
is recognized to be a special case of the Hamiltonian of 
Eq.~(\ref{hPDSsu3}), shown to have SU(3)-PDS 
of type I. As such, it has a subset of solvable eigenstates, 
Eqs.~(\ref{2N0})-(\ref{2N4k2k}),
which are members of deformed ground 
$g(K=0)$ and $\gamma^{k}(K=2k)$ bands 
with good SU(3) symmetry, $(\lambda,\mu)=(2N-4k,2k)$. 
In addition, $\hat{H}_{cri}(\beta_0=\sqrt{2})$ has the spherical states of 
Eq.~(\ref{spher}), with good U(5) symmetry, as eigenstates. 
The spherical $L=0$ state, Eq.~(\ref{nd0b0}), is 
exactly degenerate with the SU(3) ground band, Eq.~(\ref{2N0}), 
and the spherical $L=3$ state, Eq.~(\ref{nd3b0}), 
is degenerate with the SU(3) $\gamma$-band, Eq.~(\ref{2N4k2k}) 
with $k=1$. 
The remaining levels of $\hat{H}_{cri}(\beta_0=\sqrt{2})$, shown 
in Fig.~\ref{fig3}, are calculated numerically and 
their wave functions are spread over many 
U(5) and SU(3) irreps, as is evident from Fig.~\ref{fig4}. 
This situation, where some states are solvable with good U(5) symmetry, 
some are solvable with good SU(3) symmetry and all other 
states are mixed with respect to both U(5) and SU(3), 
defines a U(5) PDS of type I 
coexisting with a SU(3) PDS of type I.

\begin{figure}[bt]
\begin{minipage}{0.5\linewidth}
\centering
\rotatebox{270}{\includegraphics[width=2.3in,clip=]
{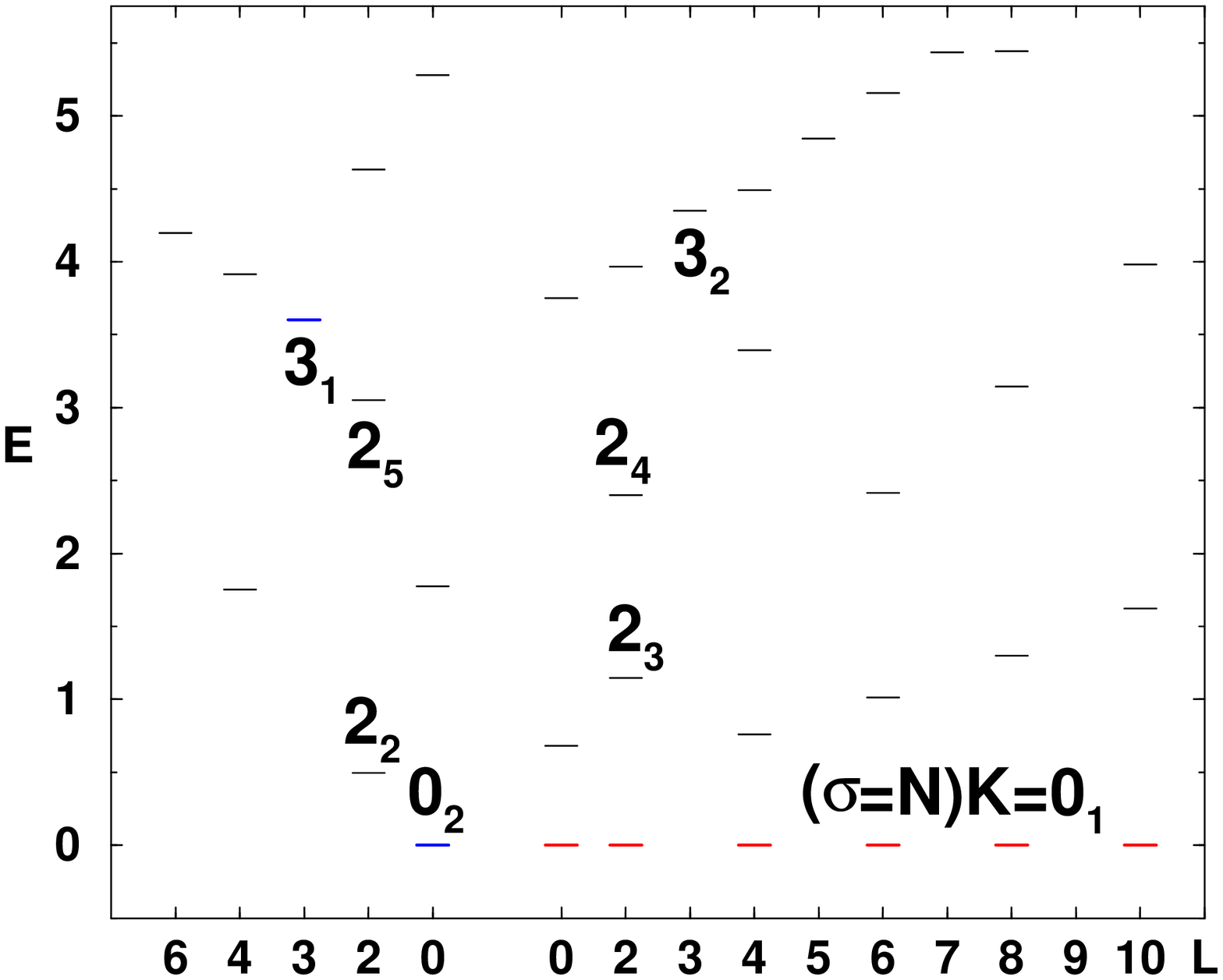}}\hspace{0.2cm}
\end{minipage}
\begin{minipage}{0.48\linewidth}
\caption{\label{fig5}
\small
Spectrum of $\hat{H}_{cri}(\beta_0=1)$, Eq.~(\ref{hcri1st}), 
with $h_2=0.05$ and $N=10$. 
$L(K=0_1)$ are the solvable states of Eq.~(\ref{deform}) 
with $\beta_0=1$, which have good 
O(6) symmetry $\sigma=N$, but broken O(5) symmetry.
$L=0_2,3_1$ are the solvable U(5) states of 
Eq.~(\ref{spher})~\cite{lev07}.}
\end{minipage}
\begin{minipage}{0.45\linewidth}
\rotatebox{270}{\includegraphics[width=2.1in,clip=]
{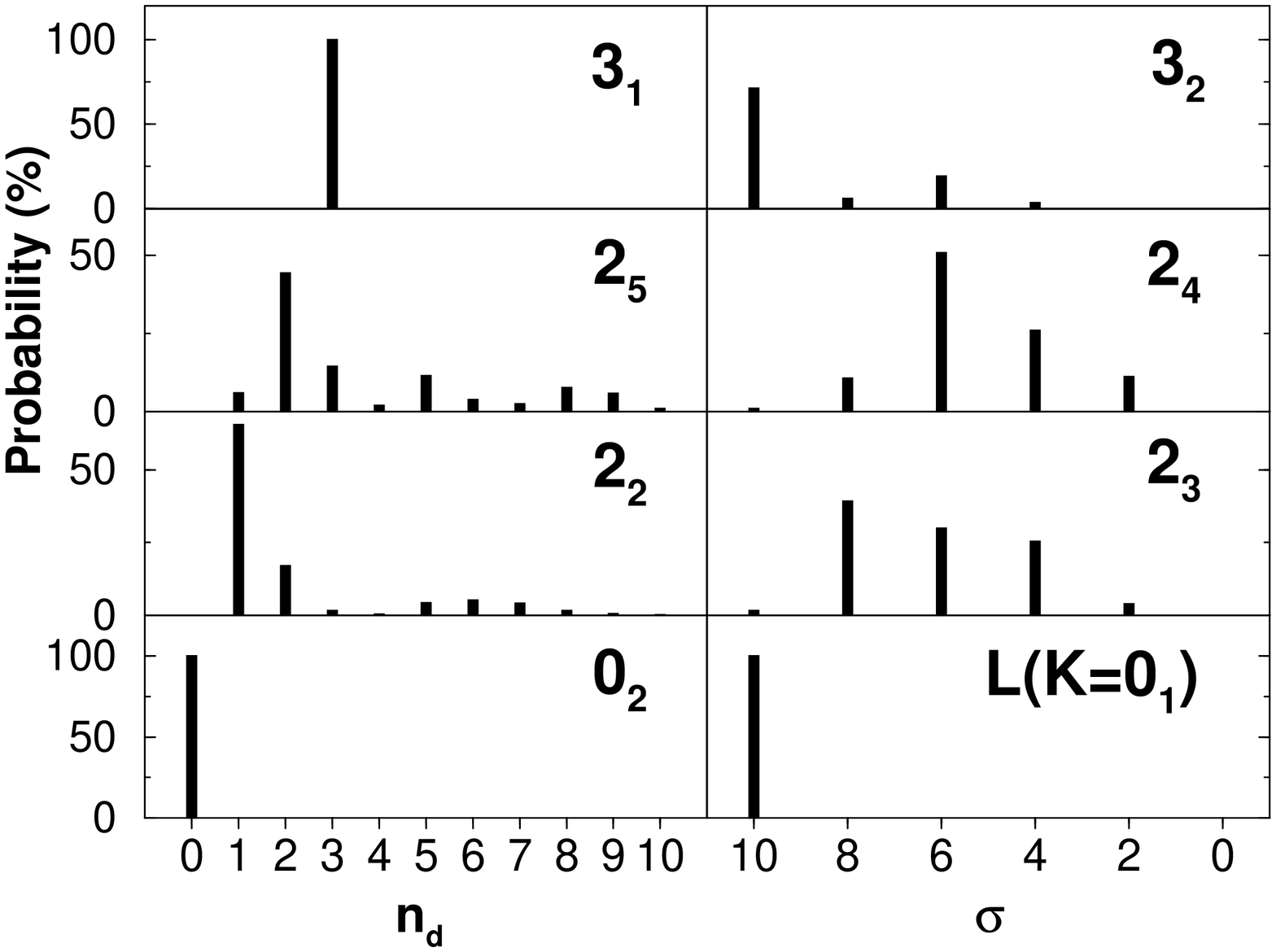}}
\caption{\label{fig6}
\small
U(5) ($n_d$) and O(6) $(\sigma)$ decomposition for selected 
spherical and deformed states in Fig.~\ref{fig5}~\cite{lev07}.}
\end{minipage}
\hspace{0.5cm}
\begin{minipage}{0.5\linewidth}
\centering
\rotatebox{270}{\includegraphics[width=2.1in,clip=]
{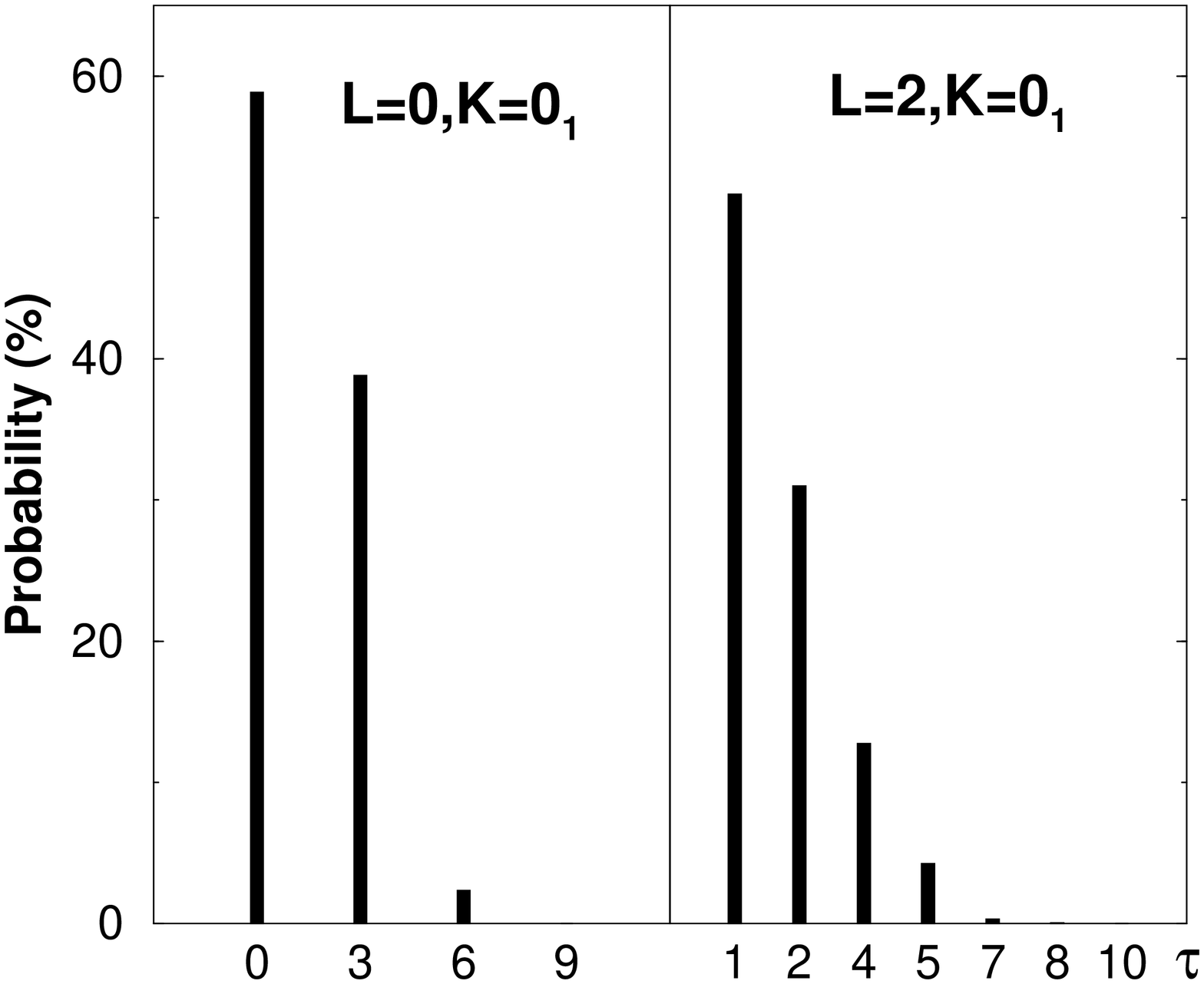}}
\caption{\label{fig7}
\small
O(5) ($\tau$) decomposition for the $L=0,2$ states, 
Eq.~(\ref{deform}) with $\beta_0=1$, members 
of the ground band ($K=0_1$) of $\hat{H}_{cri}(\beta_0=1)$.
Both states have O(6) symmetry $\sigma=N$~\cite{lev07}.}
\end{minipage}
\end{figure}

For $\beta_0=1$, the critical Hamiltonian of Eq.~(\ref{hcri1stmult}) 
involves the Casimir operators of O(5) and O(3) 
which are diagonal in the corresponding 
quantum numbers, $\sigma$ and $\tau$, 
of the O(6)-DS chain, Eq.~(\ref{chaino6}). 
It also contains 
a term involving $\hat{n}_d$ which is a scalar under O(5) 
but can connect states differing by $\Delta\sigma=0,\pm 2$ and a 
$\Pi^{(2)}\cdot U^{(2)}$ term 
which induces both O(6) and O(5) mixing
subject to $\Delta\sigma=0,\pm 2$ and $\Delta\tau=\pm 1,\pm 3$.
Although $\hat{H}_{cri}(\beta_0=1)$ is not invariant under O(6), 
it has a solvable prolate-deformed ground band, Eq.~(\ref{deform}) 
with $\beta_0=1$, which has good O(6) symmetry, 
$\langle\sigma\rangle =\langle N\rangle$, but broken O(5) symmetry. 
In addition, $\hat{H}_{cri}(\beta_0=1)$ has the spherical states of 
Eq.~(\ref{spher}), with good U(5) symmetry, as eigenstates. 
The remaining eigenstates 
in Fig.~\ref{fig5} are mixed with respect to both U(5) and O(6), 
as is evident from the decomposition shown in Fig.~\ref{fig6}. 
Apart from the solvable U(5) states 
of Eq.~(\ref{spher}), all eigenstates of $\hat{H}_{cri}(\beta_0=1)$ 
are mixed with respect to O(5) [including the solvable 
O(6) states of Eq.~(\ref{deform}) with $\beta_0=1$, 
as shown in Fig.~\ref{fig7}]. 
It follows that the Hamiltonian has a subset of states 
with good U(5) symmetry and a subset of states with good O(6) 
but broken O(5) symmetry, and all other states are mixed with respect 
to both U(5) and O(6). These are precisely the required features of 
U(5) PDS of type I coexisting with O(6) PDS of type III.

Second-order quantum phase transitions between spherical and 
deformed $\gamma$-unstable nuclei, can be accommodated in the IBM, 
by mixing the DS Hamiltonians of the U(5) chain, Eq.~(\ref{chainu5}), 
and the O(6) chain, Eq.~(\ref{chaino6}). 
Since U(5) and O(6) are incompatible, yet both chains have a common 
$O(5)\supset O(3)$ segment, we encounter a situation similar to that 
described in Eq.~(\ref{G0chains}), which gives rise to O(5)-PDS of 
type~II~\cite{Lev86}.

\section{Concluding remarks}
\label{conclusion}

The notion of partial dynamical symmetry generalizes the concepts of exact and 
dynamical symmetries. In making the transition from an exact to a 
dynamical symmetry, states which 
are degenerate in the former scheme are split but not mixed in the latter,
and the block structure of the Hamiltonian is retained.
Proceeding further to partial symmetry, some blocks or selected states in a 
block remain pure, while other states mix and lose the symmetry character. 
A partial dynamical symmetry lifts 
the remaining degeneracies, but preserves the symmetry-purity of the 
selected states.

Having at hand concrete algorithms for 
identifying and constructing Hamiltonians with PDS, is a valuable asset. 
It provides selection criteria for the a priori huge number of 
possible symmetry-breaking terms, accompanied by a rapid proliferation 
of free-parameters. This is particularly important in complicated 
environments when many degrees of freedom take part in the dynamics 
and upon inclusion of higher-order terms in the Hamiltonian. 
Futhermore, Hamiltonians with PDS break the dynamical symmetry (DS) 
but retain selected solvable eigenstates with good symmetry. The 
advantage of using interactions with a PDS is that they can be introduced, 
in a controlled manner, without destroying results previously obtained 
with a DS for a segment of the spectrum. These virtues 
greatly enhance the scope of applications of algebraic modeling 
of quantum many-body systems.

PDSs appear to be a common feature in algebraic descriptions of dynamical 
systems. They are not restricted to a specific model but can be applied 
to any quantal systems of interacting particles, bosons, as demonstrated 
in the present contribution, 
and fermions~\cite{lev11,escher00,isa08}. They are also relevant 
to the study of mixed systems with coexisting regularity and chaos,
where they lead to a suppression of chaos~\cite{walev93,levwhe96}.
 
\ack
This work is supported by the Israel Science Foundation.


\medskip
\begin{thebibliography}{99}

\bibitem{BNB}
Bohm A, N\' eeman Y and Barut A O eds 1988 
{\it Dynamical Groups and Spectrum Generating Algebras} 
(Singapore: World Scientific)

\bibitem{WIG}
Wigner E 1937 
{\it Phys. Rev.} {\bf 51} 106

\bibitem{Kerman61}
Kerman A 1961 
{\it Ann. Phys.} {\bf 12} 300

\bibitem{Elliott58}
J.P. Elliott 1958 
{\it Proc. Roy. Soc.} A {\bf 245} 128; 562 

\bibitem{Rowe85}
D.J. Rowe 1985
{\it Rep. Prog. Phys.} {\bf 48} 1419

\bibitem{GIN}
J.N. Ginocchio 1980
{\it Ann. Phys.} {\bf 126} 234

\bibitem{ibm}
Iachello F and Arima A 1987 
{\it The Interacting Boson Model} 
(Cambridge: Camb. Univ. Press)

\bibitem{ibfm}
Iachello F and Van Isacker P 1991 
{\it The Interacting Boson-Fermion Model} 
(Cambridge: Camb. Univ. Press)

\bibitem{vibron}
Iachello F and Levine R D 1994 
{\it Algebraic Theory of Molecules} 
(Oxford: Oxford Univ. Press)

\bibitem{Frank94}
A.~Frank and P.~Van~Isacker,
{\em Algebraic Methods in Molecular and Nuclear Physics},
 (New York: Wiley)

\bibitem{BIL}
Bijker R, Iachello F and Leviatan A 1984 
{\it Ann. Phys.} {\bf 236} 69; {\it ibid} 2000 
{\bf 284} (2000) 89 

\bibitem{Iachello06}
Iachello F 2006 
{\it Lie Algebras and Applications} 
(Berlin: Springer-Verlag)

\bibitem{lev11}
Leviatan A 2011, 
Prog. Part. Nucl. Phys. {\bf 66} 93 

\bibitem{gino80}
Ginocchio J N and Kirson M W 1980 
{\it Phys. Rev. Lett.} {\bf 44} 1744

\bibitem{diep80}
Dieperink A E L, Scholten O and Iachello F 1980 
{\it Phys. Rev. Lett.} {\bf 44} 1747

\bibitem{AL92}
Alhassid Y and Leviatan A 1992 
{\it J. Phys. A} {\bf 25} L1265

\bibitem{RamLevVan09}
Garc\'\i a-Ramos J E, 
Leviatan A and Van Isacker P 2009 
{\it Phys. Rev. Lett.} {\bf 102} 112502 

\bibitem{Lev86}
Leviatan A, Novoselsky A and Talmi I 1986 
{\it Phys. Lett. B} {\bf 172} 144

\bibitem{isa99} 
Van Isacker P 1999 
{\it Phys. Rev. Lett.} {\bf 83} 4269

\bibitem{levisa02}
Leviatan A and Van Isacker P 2002
{\it Phys. Rev. Lett.} {\bf 89} 222501

\bibitem{lev96}
Leviatan A 1996
{\it Phys. Rev. Lett.} {\bf 77} 818

\bibitem{Meyer85}
De Meyer H, Vanden Berghe G and Van der Jeugt J 1985
{\it J. Math. Phys.} {\bf 26} 3109

\bibitem{Berghe85}
Vanden Berghe G, De Meyer H E and Van Isacker P 1985 
{\it Phys. Rev. C} {\bf 32} 1049

\bibitem{Bona88}
Bonatsos D 1985
{\it Phys. Lett. B} {\bf 200} 1

\bibitem{Vant90}
Vanthournout J 1990
{\it Phys. Rev. C} {\bf 41} 2380

\bibitem{RDW84}
Rosensteel G, Draayer J P and Weeks K J 1984 
{\it Nucl. Phys. A} {\bf 419} 1

\bibitem{DR85}
Draayer J P and 
Rosensteel G 1985 
{\it Nucl. Phys. A} {\bf 439} 61

\bibitem{Cizewski78}
Cizewski J A  {\it et al.} 1978 
{\it Phys. Rev. Lett.} {\bf 40} 167

\bibitem{ramos00b}
Garc\'\i a-Ramos J E, Arias J M  and Van~Isacker P 2000
{\it Phys. Rev. C} {\bf 62} 064309 

\bibitem{lev07}
Leviatan A 2007
{\it Phys. Rev. Lett.} {\bf 98} 242502

\bibitem{carr10} 
Carr L Ed. 2010
{\it Understanding Quantum Phase Transitions} 
(CRC press)

\bibitem{escher00}
Escher J and Leviatan A 2000
{\it Phys. Rev. Lett.} {\bf 84} 1866; 
{\it ibid} 2002 
{\it Phys. Rev. C} {\bf 65} 054309 

\bibitem{isa08} 
Van Isacker P and Heinze S 2008 
{\it Phys. Rev. Lett.} {\bf 100} 052501

\bibitem{walev93}
Whelan N , Alhassid Y and Leviatan A 1993 
{\it Phys. Rev. Lett.} {\bf 71} 2208

\bibitem{levwhe96}
Leviatan A and Whelan N D 1996
{\it Phys. Rev. Lett.} {\bf 77} 5202

\end{thebibliography}
\end{document}